# Inorganic Graphenylene: A Porous Two-Dimensional Material With Tunable Band Gap

*E. Perim1, R. Paupitz*2, P. A. S. Autreto1 and D. S. Galvao1.*

1Instituto de Física 'Gleb Wataghin', Universidade Estadual de Campinas, 13083-970, Campinas, São Paulo, Brazil.

2Departamento de Física, IGCE, Universidade Estadual Paulista, UNESP, 13506-900, Rio Claro, SP, Brazil.

*Corresponding Author: Email: paupitz@rc.unesp.br ; +55 19 3526 9156

ABSTRACT: By means of ab initio calculations we investigate the possibility of existence of a boron nitride (BN) porous two-dimensional nanosheet which is geometrically similar to the carbon allotrope known as biphenylene carbon. The proposed structure, which we called Inorganic Graphenylene (IGP), is formed spontaneously after selective dehydrogenation of the porous Boron Nitride (BN) structure proposed by Ding et al. We study the structural and electronic properties of both porous BN and IGP and it is shown that, by selective substitution of B and N atoms with carbon atoms in these structures, the band gap can be significantly reduced, changing their behavior from insulators to semiconductors, thus opening the possibility of band gap engineering for this class of two-dimensional materials.

Keywords: Graphene, hexagonal Boron Nitride, Simulation, Nanotechnology.

**Introduction**

Carbon based nanostructures have been object of intense investigations since the discovery of fullerenes[1] and nanotubes[2]. There is a renewed interest in these structures given by the recent isolation of graphene[3]. Graphene and of several of its derivatives[4,5], present a great number of interesting physical properties. Many of the proposed technological applications are based on the outstanding mechanical and electronic properties of graphene which, for instance, presents the highest known Young modulus for any material and exceptional carrier mobility[6]. Despite this, graphene-based electronics is nowadays hindered by the fact that graphene is a zero-gap semiconductor, leading to the pursuit of efficient methods for opening its band gap. Many methods have been tried in order to control its band gap, including chemical methods like hydrogenation[7,8,9], oxidation[10] and/or fluorination[11,12] or the production of graphene nanoribbons (GNRs)[13]. Although these methods have been partially successful, graphene electronic properties are highly sensitive to the presence of structural defects, resulting in a strong dependency on controlled synthesis techniques[14]. In this context, many other two dimensional structures have become objects of interest, from inorganic structures, such as silicene[15], germanene[16] and hexagonal boron nitride[17], to organic ones, such as porous graphene[18] and biphenylene carbon (BPC)[19]. Both porous graphene and BPC are considered promising nanostructures for nanotechnology applications, since they have the advantage of retaining some of graphene properties while being intrinsically non-zero gap structures. Most surprisingly, BPC had already been predicted more than 40 years ago[20]. Recent advances have made possible the synthesis of small fragments of porous graphene[21] but the BPC synthesis remains elusive.

Hexagonal boron nitride (hBN) is also known as the inorganic analogue of graphene, presenting the same honeycomb lattice configuration, which can be seen in Figure 1(a). It has almost identical bond lengths and cell parameters. As a consequence, hBN can generate many nanostructures equivalent to the carbon ones, from closed cage structures[22] and nanotubes[23] to the recently synthesized nanoscrolls[24,25]. Despite these similarities, the lower symmetry of the hBN lattice and the chemical and physical dissimilarities between B and N elements, lead to very distinct electronic properties of BN nanostructures. Taking these aspects into consideration, we carried out a study about the electronic properties and stability of both porous BN[26] and a possible analogue of BPC, which we named Inorganic Graphenylene (IGP) composed of boron and nitrogen atoms, shown in Figure 1(b) and Figure 1(c) respectively. Based on ab initio calculations, we show that a selective removal of hydrogen atoms from a porous BN sheet triggers a process of spontaneous conversion of the resulting structure to a stable IGP sheet, which happens to be stable. Taking into account the great interest that emerged in the past few years on the properties of B-N-C materials[27], it is worth the effort to make an investigation about possible physical properties of B-N-C structures with geometry similar to BPC and IGP. For that reason, we included in this study an analysis about the possibility of band structure modifications and band gap engineering made through atomic substitution of B and N atoms by carbon atoms in porous BN and IGP. Despite the fact that both these pure BN sheets have large band gaps it was possible to show that, via substitution with carbon atoms, we can effectively tune the gap values, reaching values below 1eV. We also present some structural aspects of each structure.

**Computational Methods**

First principle calculations were performed within the Density Functional Theory (DFT) methodology using Dmol3[28] package as implemented on the Accelrys Materials Studio software. We adopted the Perdew-Burke-Ernzerhof (PBE) functional under the generalized gradient approximation (GGA). A 2x2 supercell was used in all calculations with sheet separation of 25Å to prevent inter-sheet interactions. The geometry optimizations were carried out with all atoms free to move and full cell optimization. The convergence criteria were of $1.10^{-6}$Ha in energy, $5.0 \cdot 10^{-4}$ Ha/Å in maximum force and 0.005Å in maximum displacement. During geometry optimizations we used a k-point sampling of 7x7x1, yielding a k-point separation of ~0.01 Å$^{-1}$ in the reciprocal space at the periodic directions, while band structures were calculated using a k-point sampling of 15x15x1, yielding a k-point separation of ~0.006 Å$^{-1}$. Core electrons were treated explicitly and a double numerical plus polarization (DNP) basis set was used. Dynamics were carried out using an NVT ensemble with a 1.0fs time step and a massive generalized Gaussian moments (GGM) thermostat.

**Results and Discussion**

The porous BN structure has four different bonds, a B-H bond, a N-H bond and two different B-N bonds, one as part of the rings and one linking different rings. Our obtained results are 1.204 Å for the B-H bond, 1.014 Å for the N-H bond, 1.438 Å for the B-N bonds inside one ring and 1.495 Å for the B-N bonds linking different rings. The unit cell parameters were 7.65 Å for *a* and *b*, 90º for α and β and 120º for γ. hBN and porous BN structure can be seen in Figure 1(a) and (b), respectively. These results are in very good agreement with previous[19] work. The band structure of the porous BN, as seen in Figure 1(b), shows a wide band gap of 4.57eV which is also in very good accordance with previous work[19]. A significant density of states was observed below the Fermi energy, but for energies slightly above this value no accessible states were found, corroborating the insulating character of the material.

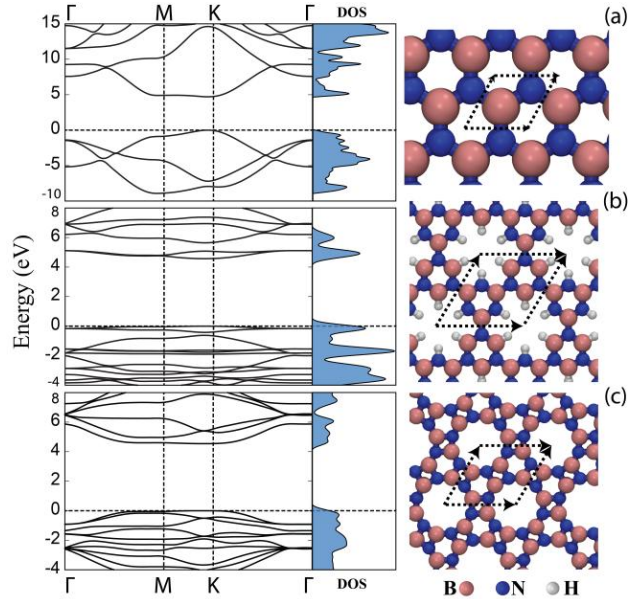

**Figure 1.** Structural models and the corresponding band structure and Density of States (DOS) for the structures considered in this work. (a) hBN, with two atoms in the unit cell, (b) porous BN with 18 atoms in the unit cell, and (c) Inorganic Graphenylene (IGP) with 12 atoms in the unit cell. Dark (blue online) spheres represent nitrogen atoms while boron ones are indicated by the light (pink online) colors. Unit cell is highlighted by dashed lines in each case.

If we remove the hydrogen atoms from the optimized porous BN structure and let it freely evolve in time, the nanosheet undergoes a structural change (see Supplementary Information), forming alternating hexagonal and rhombus-like rings with large pores, as seen in Figure 1(c). This structure is extremely similar in shape to an organic one previously reported[19], although in this case the four-member rings are not squares. Thus we named it the inorganic graphenylene (IGP). The optimized structure has three different B-N bond distances, depending on the specific pair location. For instance, if in a four-member ring, in a six-member ring or both. These values are respectively 1.461 Å, 1.408 Å and 1.503 Å. In rhombus-like rings, the nitrogen centered angles are of 83.7º, while the boron centered ones are of 96.3º. Lattice constants are calculated to be 6.8 Å for both *a* and *b,* with hexagonal symmetry. The IGP band structure, shown in Figure 1(c), presents a band gap of 4.14 eV, value similar to that obtained for porous BN. There are also some similarities between their density of states, shown in Figure 1(b) and 1(c), both presenting many accessible states below the Fermi level but no accessible states for values slightly above that energy. Regarding the physical properties determined by electronic band structure, IGP and BPC electronic are quite different, since the former is a large gap insulator while the latter is a semi-conductor.

It would be of great interest if we could perform band gap engineering on this class of materials. Our calculations show that this can effectively be done by selective substitution, on these sheets, of boron and nitrogen by carbon atoms. To explore this possibility, we calculated the band structure of several porous BN-like and IGP-like structures in which up to four BN pairs were substituted by carbon atoms. All substitutions were done combining each B to C substitution with an N to C one. Such procedure is convenient for our purposes since in this way we keep the total number of electrons constant and the same of the original BPC or porous BN. By these substitutions, which represented changes of less than 20% of the supercell atoms, we were able to reduce the band gap value by one order of magnitude, as we discuss in the following.

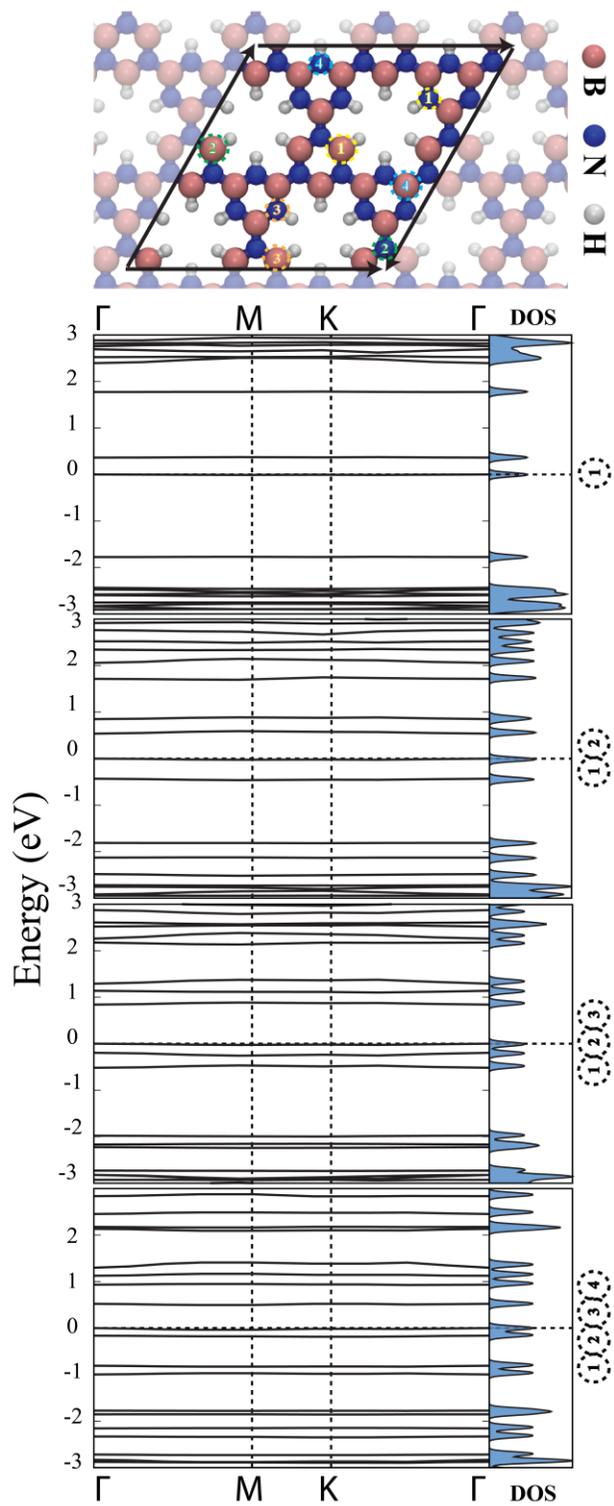

**Figure 2.** B-N by C substitutions considered for Porous BN calculations using 2x2 supercells. The label numbering indicates the position of the substitutions. The corresponding band structure for each case is also presented.

In the case of porous BN we used supercells, as defined in Figure 2, to make the BN by C substitutions and to carry out band structure calculations. From one up to four substituted pairs of BN, the band gap value varies from 4.57 eV to 0.49 eV, as one can see in Table 1. Comparing Figure 1(b) and Figure 2 one can see that the number of accessible electronic states above the Fermi level increases for the carbon substituted structures, effectively converting the former insulator into a semiconductor.

**Table 1.** Lattice angles, lattice parameters, cohesive energy ($E_c$) and bandgap values for the considered structures.

| Structure | # of carbons per supercell | α | β | γ | a (Å) | b (Å) | $E_c$(eV/atom) | bandgap (eV) |
|---|---|---|---|---|---|---|---|---|
| Porous BN | 0 | 90° | 90° | 120° | 15.30 | 15.30 | 5.34 | 4.57 |
| | 2 | 89.6° | 90.9° | 119.9° | 15.31 | 15.31 | 5.31 | 0.35 |
| | 4 | 90.1° | 90.8° | 119.7° | 15.31 | 15.21 | 5.28 | 0.54 |
| | 6 | 88.8° | 87.7° | 119.1° | 15.33 | 15.13 | 5.28 | 0.82 |
| | 8 | 88.3° | 87.7° | 119.3° | 15.34 | 15.22 | 5.26 | 0.49 |
| IGP | 0 | 90° | 90° | 120° | 13.68 | 13.68 | 6.56 | 4.14 |
| | 2 | 88.9° | 86.8° | 120.0° | 13.67 | 13.73 | 6.50 | 0.82 |
| | 4 | 88.6° | 87.9° | 119.5 | 13.56 | 13.62 | 6.47 | 1.12 |
| | 6 | 88.2° | 90.8° | 120.0° | 13.71 | 13.73 | 6.43 | 1.01 |
| | 8 | 88.1° | 89.0° | 120.1° | 13.69 | 13.79 | 6.36 | 0.05 |

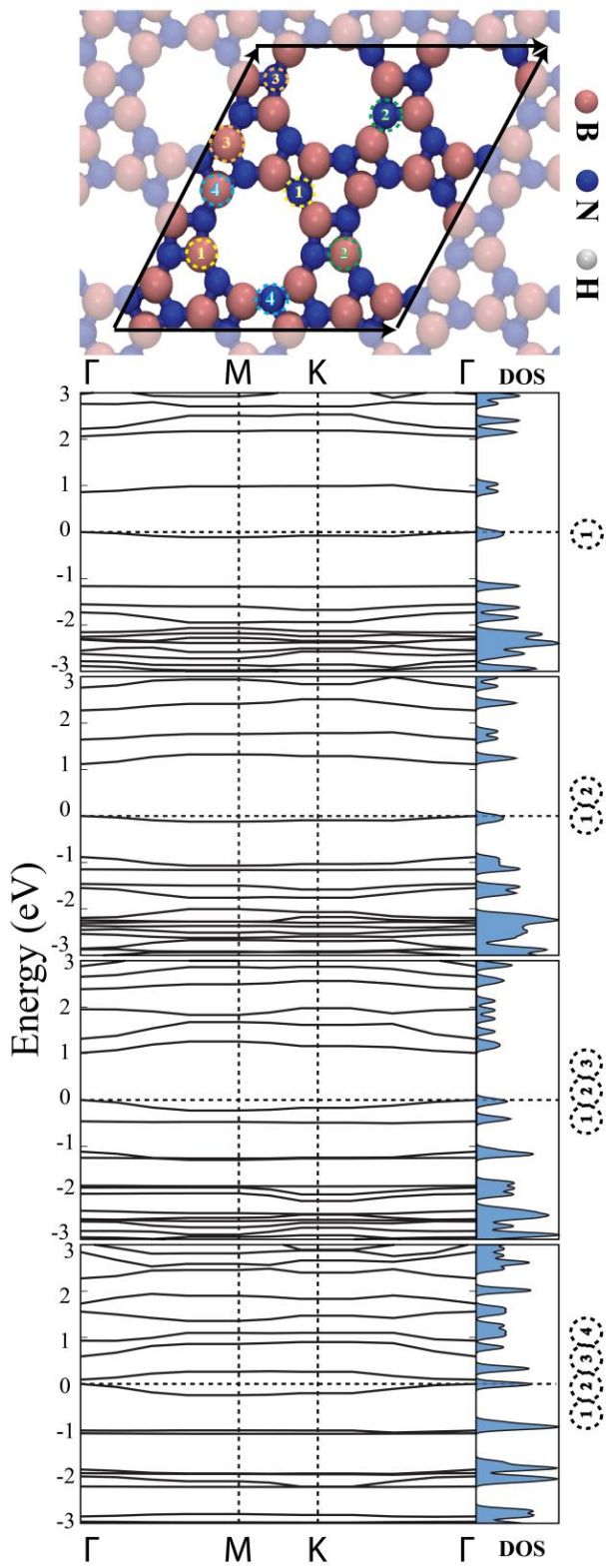

**Figure 3.** B-N by C substitutions considered for IGP calculations using 2x2 supercells. The label numbering indicates the position of the substitutions. The corresponding band structure for each case is also presented.

In the case of IGP, we also used supercells, as can be seen in Figure 3. The atom substitutions decreased the band gap from 4.14 eV to as low as 0.08 eV and once more, as shown in Figure 3, we notice the appearance of accessible electronic states above the Fermi level. A direct comparison between the gap tuning obtained by carbon substitutions in porous BN and IGP can be seen on Figure 4.

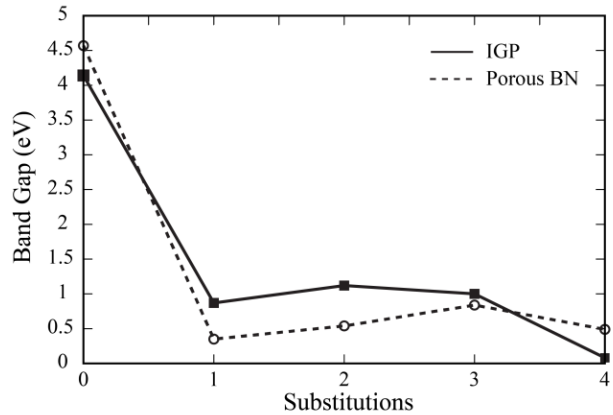

**Figure 4.** Band gap values versus the number of carbon substitutions. Each substitution corresponds to a BN pair being replaced by two carbons.

To have a measure of the stability of each structure we calculated the cohesive energy, which represents the energy that would be required to decompose the structure into isolated atoms. This quantity is defined as the difference between the sum of the energies of the isolated atoms and the energy of the structure as in $E_c = \sum n_x E_x - E_T$, where $n_x$ stands for the number of atoms of type X in the structure, $E_x$ is the energy of an isolated atom of type X and $E_T$ is total energy of the structure. The results are shown in Table 1 and indicate that the porous BN structures are less stable than their IGP counterparts by an energy of over 1eV/atom and also that by adding carbon atoms a slight decrease on the stability is caused (of no more than 3%) on both types of structures. The higher degree of stability presented by the IGP structures could suggest a possible approach of using an electron beam as a mean of selectively removing hydrogen atoms by knock-on effect[29] from a pBN sheet and thus causing the spontaneous conversion to an IGP sheet. The chemical potential for removing one hydrogen atom bound to a nitrogen one was estimated as ~4eV.

The small decrease on the cohesive energy of less than 1% per substitution (as seen in Table 1) suggests that these substitutions could be experimentally feasible. Also, we can compare the cohesive energies for the pBN and the IGP with those of hexagonal BN (hBN) and zincblend BN (zcBN), which are stable phases already successfully produced. The most stable phase is hBN with AB packing, presenting a cohesive energy of 6.997eV/atom, followed by hBN with AA packing, which presents a cohesive energy of 6.992eV/atom. zcBN presents a cohesive energy value of 6.953eV/atom. By comparison, we see that the IGP presents a cohesive energy ~5% lower than that of zcBN and ~6% lower than that of hBN. These can be considered promising results towards the possibility of experimentally producing IGP structures.

Another interesting effect caused by these substitutions is related to the frontier orbitals configurations, namely the HOCO (Highest Occupied Crystal Orbital) and LUCO (Lowest Unoccupied Crystal Orbital), whose degree of delocalization is related to electronic conduction behavior of the material. These changes in frontier orbitals for all the structures considered in this work can be seen in Supplementary Material in Figures S1-S10.

These are very promising results which suggest these structures could be used in a wide array of electronic devices since their band gap values can be tuned in a very wide range by carbon substitutions. The study of the mechanical properties of these novel structures could also reveal interesting aspects and will be object of future works.

**Summary and Conclusions**

In summary, we investigated distinct aspects of two classes of inorganic two-dimensional nanostructures, porous BN and inorganic graphenylene. It was shown that by selective removal of hydrogens from porous BN, it is possible to induce a spontaneous structural change, forming a novel and not yet reported structure, which we named inorganic graphenylene. Both these structures present strong insulating behavior and our simulations indicate they are stable at room temperature. Additionally, we demonstrated that the band gap of these two materials can be tuned via atomic substitution of BN pairs by carbon atoms, which change them from insulating to semi conducting materials. Taking into account our results, we believe these structures can be very promising to many technological applications, leading to the design of novel nano electronic devices. Their mechanical and transport properties should be further investigated in order to fully understand the technological potential of this class of materials.

While finalizing this article we became aware of a very interesting work studying structures similar to our carbon substituted porous BN[30], which the authors called B-C-N hybrid porous sheet, and showing the shift on the light absorption towards visible wavelengths. This just reinforces the interest in this kind of structure and opens new perspectives to novel applications.

ACKNOWLEDGEMENTS: The authors would like to thank Prof. van Duin, Dr. Luiz H. G. Tizei and Dr. Vitor T. A. Oiko for helpful discussions for helpful discussions and Prof.

Alexandru T. Balaban for a critical reading of the manuscript. This work was supported in part by the Brazilian Agencies CNPq and FAPESP. The authors thank the Center for Computational Engineering and Sciences at Unicamp for financial support through the FAPESP/CEPID Grant # 2013/08293-7. RP also acknowledges FAPESP Grant # 2011/17253-3.

**Supporting Information Available:** Frontier orbitals figures and molecular dynamics videos. This material is available free of charge via the Internet at http://pubs.acs.org.

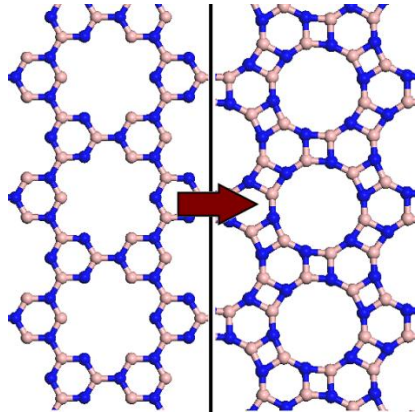